\documentclass[osajnl,twocolumn,showpacs,superscriptaddress,10pt]{revtex4-1} 
\usepackage{amsmath,amssymb,graphicx}
\begin{document}
\newcommand{\note}[1]{\marginpar{\raggedright \tiny{#1}}}
\newcommand{\snote}[1]{} 
\renewcommand{\note}[1]{}
\reversemarginpar
%
\newcommand{\bnote}[1]{
\hfill
\framebox[0.6\textwidth]{
\begin{minipage}[t]{0.5\textwidth}{
\setlength{\parskip}{2ex}
\raggedright #1
}
\end{minipage} }\vspace*{2ex} }
\setlength{\unitlength}{1cm} 
\newlength{\xyfigsize}
\setlength{\xyfigsize}{1.0 cm}
\newlength{\xydfigsize}
\setlength{\xyfigsize}{15.0 cm}
\newcommand{\intthree}{\int \!\!\! \int \!\!\! \int}
\newcommand{\inttwo}{\int \!\!\! \int}
\newcommand{\be}{\begin{equation}}
\newcommand{\ee}{\end{equation}}
\newcommand{\benn}{\begin{displaymath}}
\newcommand{\eenn}{\end{displaymath}}
\newcommand{\ba}{\begin{eqnarray}}
\newcommand{\ea}{\end{eqnarray}}
\newcommand{\nn}{\nonumber}
\newcommand{\ds}{\displaystyle}
\newcommand{\hf}{\hfill}
\renewcommand{\vec}[1]{\mbox{\boldmath $#1$}}
\newcommand{\mat}[1]{\mbox{\bf #1}}
\renewcommand{\d}[1]{{\rm d}#1}
\newcommand{\e}[1]{{\rm e}^{#1}}
\newcommand{\half}{\mbox{$\frac{1}{2}$}}
\newcommand{\sfrac}[2]{\mbox{$\frac{#1}{#2}$}}
\newcommand{\rb}{Rayleigh--Brillouin}
\newcommand{\RB}{Rayleigh--Brillouin}
\newcommand{\rbs}{Rayleigh--Brillouin scattering}
\newcommand{\Rbs}{Rayleigh--Brillouin scattering}
\newcommand{\Srbs}{Spontaneous Rayleigh--Brillouin scattering}
\newcommand{\Srb}{Spontaneous Rayleigh--Brillouin}
\newcommand{\srb}{spontaneous Rayleigh--Brillouin}
\newcommand{\srbs}{spontaneous Rayleigh--Brillouin scattering}
\newcommand{\Crbs}{Coherent Rayleigh--Brillouin scattering}
\newcommand{\crbs}{coherent Rayleigh--Brillouin scattering}
\newcommand{\crb}{coherent Rayleigh--Brillouin}
\newcommand{\wt}[1]{\widetilde{#1}}
\newcommand{\pr}[1]{p = #1 \; {\rm bar}}
\newcommand{\etab}{\eta_{\rm b}}
\newcommand{\eb}[1]{\eta_{\rm b} = #1 \times 10^{-5} \: {\rm kg m}^{-1}
{\rm s}^{-1}}
\newcommand{\seb}[1]{\sigma_{\eta_{\rm b}} = #1 \times 10^{-5} \: {\rm kg \:
m}^{-1} {\rm s}^{-1}}
\newcommand{\eberr}[2]{\sigma_{\eta_{\rm b}} = #1 \pm #2 \times 10^{-5}
\: {\rm kg \: m}^{-1} {\rm s}^{-1}}
%

\title{Rayleigh--Brillouin scattering of carbon dioxide}

\author{Z. Y. Gu}
\affiliation{Department of Physics and Astronomy, LaserLaB, VU
University, De Boelelaan 1081, 1081 HV Amsterdam, The Netherlands}

\author{W. Ubachs}\email{Corresponding author: w.m.g.ubachs@vu.nl}
\affiliation{Department of Physics and Astronomy, LaserLaB, VU
University, De Boelelaan 1081, 1081 HV Amsterdam, The Netherlands}

\author{W. van de Water}
\affiliation{Physics Department, Eindhoven University of Technology,
Postbus 513, 5600 MB Eindhoven, The Netherlands}

\begin{abstract}
The spectral lineshape of \srbs\ in CO$_2$ is studied in a range of
pressures.  The spectrum is influenced by the bulk viscosity
$\eta_b$, which is a relaxation phenomenon involving the internal
degrees of freedom of the molecule.  The associated relaxation rates
can be compared to the frequency shift of the scattered light, which
demands precise measurements of the spectral lineshape.  We find
$\eta_b = (5.7 \pm 0.6) \times 10^{-6} \: {\rm kg \: m}^{-1} {\rm
s}^{-1}$ for the range of pressures $p= 2-4$ bar and for room temperature conditions.
\end{abstract}

\ocis{(290.5830) Scattering, Brillouin; (290.5840) Scattering, molecules; (290.5870)  Scattering, Rayleigh; (010.0010) Atmospheric and oceanic optics;  (140.6810)  Thermal effects; .}

\maketitle 

We present a study on the precise spectral shape
of light which is quasi--elastically scattered off a gas of CO$_2$
molecules. The spectrum depends on the internal degrees of freedom of
the molecule.  For CO$_2$ the rotational relaxation rate, the rate of energy exchange between the translational and rotational degrees of freedom through collisions, is
comparable to the frequency shift of the scattered light, while the
vibrational relaxation rates are comparable to the MHz frequencies of
conventional ultrasound experiments.  This makes an interesting case
to test models of the spectral lineshape using precise experiments.
The relaxation of internal degrees of freedom determines the bulk
viscosity $\eta_b$, which is a parameter in models of the line shape.

This study was inspired by a recent debate in the literature
about the precise value of $\eta_b$ of CO$_2$ \cite{pan.co2}.  It was
made possible by the construction of a new experimental setup which
provides spectra with unprecedented statistical accuracy
\cite{vu.rsi}.
Accurate information about the spectral line shape of Rayleigh-Brillouin backscattered light is of relevance for remote sensing applications in the Earth atmosphere~\cite{Witschas} as well as for oceanographic applications~\cite{Liang}. In particular detailed information is needed for ESA's future ADM-Aeolus mission which will provide global
observations of wind profiles from space utilizing an active
satelite-based remote sensing instrument in the form of a Doppler
Wind Lidar \cite{ESA-AMD}. Information on scattering from CO$_2$ is of relevance
for investigations of the atmospheres of Venus and Mars where carbon dioxide is the
main constituent.

Rayleigh-Brillouin scattering is caused by spontaneous density
fluctuations: sound.  The spectral lineshape of the scattered light
is influenced by the damping of sound through the molecular viscosity of
the gas.  If the gas consists of molecules with internal degrees of
freedom, such as rotation or vibration, the viscosity is also
influenced by the relaxation of these freedoms.  For CO$_2$ at
atmospheric pressures, the relaxation time for rotational motion
$\tau_r$ is $\tau_r = 3.8\times10^{-10}\; {\rm s}$, while for
vibrational motion it is $\tau_v = 6\times 10^{-6}\; {\rm s}$
\cite{Lambert1977}.  The relaxation of internal modes of motion
determines the bulk viscosity $\eta_b$.  At low sound frequencies $f$
in the range of MHz, both rotational and vibrational modes couple
with translation, and $\eta_b$ is large, $\eta_b = 1.46 \times
10^{-2}\: {\rm kg \: m}^{-1} {\rm s}^{-1}$, however, in light
scattering experiments, the typical period of sound is ${\cal
O}(10^{-9}\; {\rm s})$, which is much shorter than $\tau_v$, so that
the vibrational modes are frozen, and the bulk viscosity is reduced
dramatically.
This was noticed by Lao {\it et al.} \cite{lao.1976a}, and again by
Pan {\it et al.} \cite{pan.co2} in the context of \crbs. Lao {\it et
al.} find $\eta_b = 4.6\times 10^{-6}$~kg~m$^{-1}$s$^{-1}$, while Pan
{\it et al.} find $\eta_b = 3.7\times 10^{-6}$~kg~m$^{-1}$s$^{-1}$.
In \crbs, density variations are induced by dipole forces by crossing
laser beams. \Srbs\ and \crbs\ share the same statistical description
of the scattered light spectrum, and both can be used to determine
the bulk viscosity at hypersound frequencies.  However, coherent
scattering results in a large increase of the scattered light
intensity \cite{pan.pra,ru.jcp.2010}.
Note also that the scattering spectral lines shapes in the coherent
and spontaneous forms of Rayleigh-Brillouin scattering are markedly
different, therewith providing independent means to determine the
bulk viscosity of a gas.

Our \srbs\ experiments are in the kinetic regime, where the inverse
scattering wavevector can be compared to the mean free path between
collisions.  This regime is characterized by a non--uniformity
parameter, $y = {\cal O}(1)$, where $y$ is defined as $ y = {p} / {(k
\: v_{0} \: \eta)} = {n k_\mathrm{B} \: T}/{(k \: v_{0} \: \eta)},
$ with $k$ the scattering wave vector, $n$ the number density,
$k_\mathrm{B}$ the Boltzmann constant, $T$ the temperature, $p$ the
pressure, $v_{0}$ the thermal velocity, $v_{0} = (2 \: k_\mathrm{B}
\: T / M)^{1/2}$, with $M$ the mass of the molecular scatterers, and
$\eta$ the (shear) viscosity.
Accordingly, the Boltzmann equation must be used to describe the
scattered light spectrum.  In the Tenti model the collision integral
of the linearized Boltzmann equation is approximated with the Wang
Chang and Uhlenbeck approach, using six (S6) or seven (S7) moments
\cite{tenti1,tenti2}.

We revisit the bulk viscosity of CO$_2$ and present scattered light
spectra of CO$_2$ for pressures $p = 1-4$ bar.  In
order to extract the bulk viscosity from the spectrum, the measured
\rbs\ spectra are compared to spectra predicted using the Tenti
models, with $\eta_b$ used as a fit parameter.  Since the bulk
viscosity is a relaxation parameter, and since the mean free time
between collisions is inversely proportional to pressure, we expect a
(slight) dependence of $\eta_b$ on pressure.

From a straightforward extension of the arguments
in~\cite{Chapman1970}, the frequency--dependent bulk viscosity can be
related to the relaxation time of the internal degrees of freedom,
\be
   \eta_{\mathrm{b}} = 2 n k_\mathrm{B} T
   \left| \frac{\sum_j N_j \tau_j (1 + i \omega \tau_j)^{-1}}
   {N \left( 3 + \sum_j N_j (1 + i \omega \tau_j)^{-1} \right)}
   \right|
\label{eq:relaxation}
\ee
where $N_j$ is the number of internal degrees of freedom with
relaxation time $\tau_j$, $N = 3 + \sum_j N_j$ is the total number of
degrees of freedom, $n$ the molecular number density, and where it is
assumed that the internal degrees of freedom do not interact with
other ones having a different relaxation time, and the density is
small.
When the frequency of sound waves $\omega = 2\pi f$ is much larger
than $1/\tau_j$, the mode $j$ remains frozen and does not contribute
to the bulk viscosity.  On the other hand, when $\omega$ is much
smaller than all relaxation rates, Eq.\ (\ref{eq:relaxation}) reduces
to the relation \cite{Chapman1970}
\be
\etab = 2 n k_\mathrm{B} T \sum_j N_j \tau_j / N^2.
\ee
Since an increase of the pressure results in an increase of the collision rates thus a decrease of the relaxation time of internal modes of motion, and since the sound frequency and the relaxation time appear in the combination $\omega\tau_j$, we expect that the bulk viscosity {\em increases} with increasing pressure.


The interpretation of $\eta_b$ as a relaxation parameter is not
without controversy \cite{meador95}.  Here we use $\eta_b$ as a
parameter in a kinetic model for the scattered light line shape.  In
the context of a continuum description, Meador {\it et al.}
\cite{meador95} also arrive at an equation expressing $\eta_b$ in
terms of a relaxation time, which they deem incomplete.  Similarly,
also Eq.\ (\ref{eq:relaxation}) cannot be complete as it still
contains the multiplicity $N_j$ of frozen modes for which $\omega \:
\tau_j = \infty$.

\begin{figure}[]
\centerline{\includegraphics[width=.8\columnwidth]{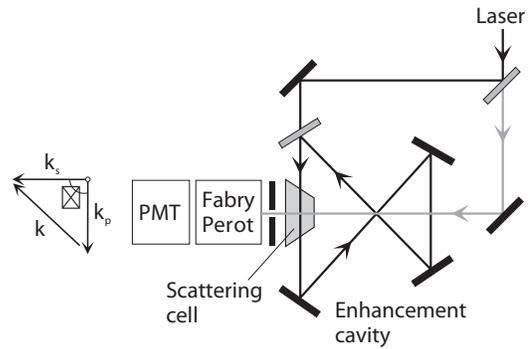}}
\caption{Schematic diagram of the experimental setup for \srbs\ (not to
   scale).  The UV laser beam (full black line) is reflected several
   times in the enhancement cavity to increase the scattering
   intensity. A reference beam (gray line), split off the main beam,
   is used for detector alignment.  Scattered light is detected at
   $90^\circ$ using a pinhole, a Fabry-Perot interferometer and a
   photo-multiplier (PMT).}\label{fig.s.setup}
\end{figure}

A schematic view of the used setup used for \srbs\ is shown in Fig.\
\ref{fig.s.setup}, with a detailed description provided in
\cite{vu.rsi}.  Briefly, the light from a narrowband continuous--wave
laser is scattered off the CO$_2$ gas contained in a
temperature-controlled gas cell.
The laser is a frequency-doubled Ti:Sa laser delivering light at
366.8~nm, 2~MHz bandwidth and 500~mW of output power.  The long-term
frequency drift was measured with a wavelength meter to be smaller
than 10~MHz per hour.
The scattered light is collected at an angle of $90^\circ$ from an
auxiliary focus inside an enhancement cavity, in which a
scattering-cell is mounted. The cell is sealed with Brewster windows.
The circulation of the light inside the enhancement cavity amplifies
the power by a factor of 10.
A hemispherical scanning Fabry-Perot interferometer (FPI) is used to
resolve the frequency spectrum of the scattered light.  The drift of
the laser frequency is much smaller than the drift of the FPI, with both drifts being corrected for by a frequency
linearization procedure. All experiments were performed in CO$_2$ gas at $T=296.5 \pm 0.5 $K.

\begin{figure*}[]
\centerline{\includegraphics{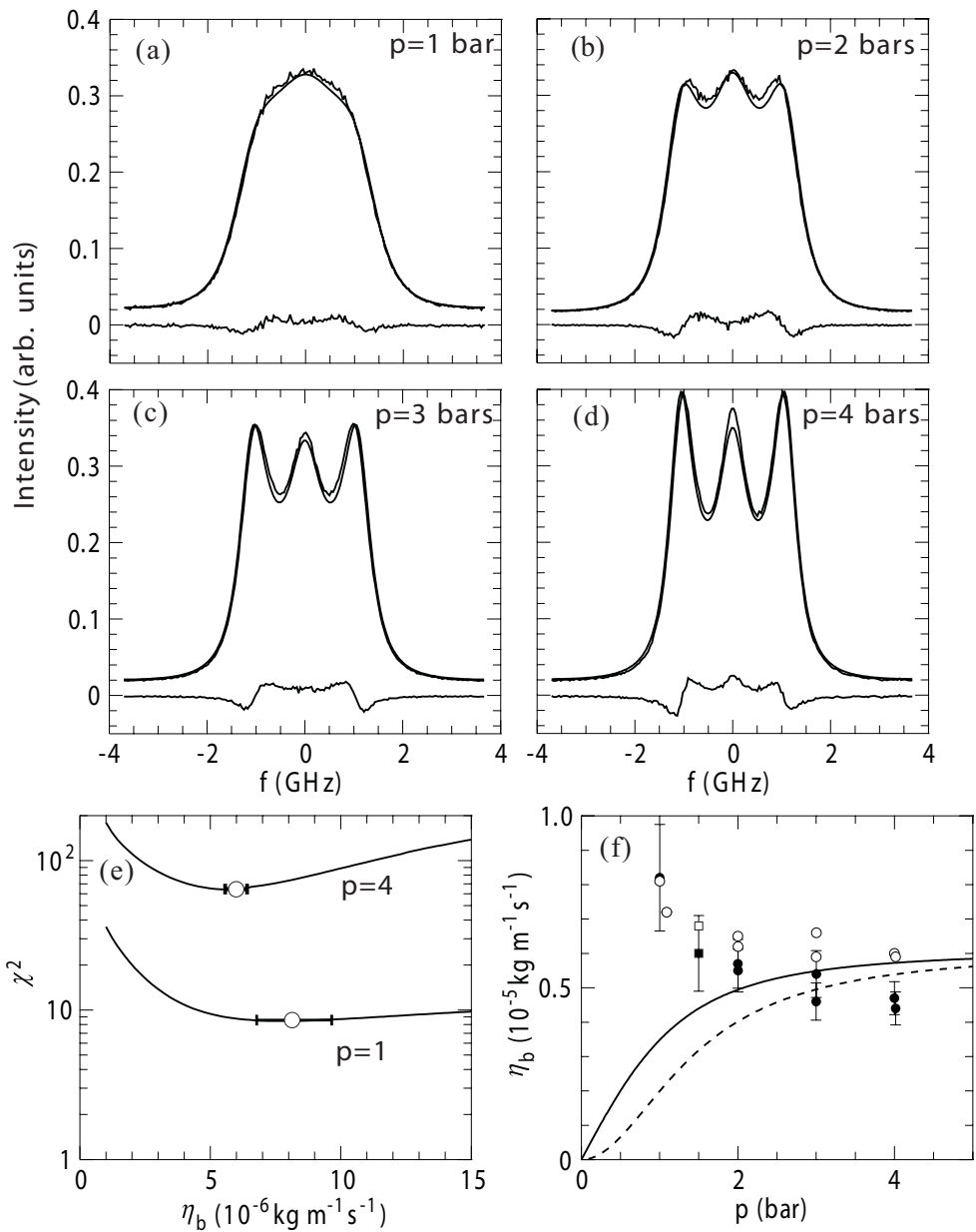}}
\caption{(a-d) CO$_2$ \rbs\ spectra at pressures $p = 1, 2, 3$ and 4 bars and for conditions of $296.5\pm0.5$ K.  The spectra are shown together with the Tenti S7
model; the lower line indicates the difference between the model and
experiment. The model calculations include a convolution with the instrument
function of the FP-analyzer. The used bulk viscosities are indicated by the open
symbols in frame (f).
(e) Lines are $\chi^2$ differences between the experimental spectra
and the Tenti S7 model as a function of $\eta_b$, the open balls
indicate the minimum $\chi^2$.
(f) Symbols indicate the bulk viscosity obtained by fitting the Tenti
model to the experimental spectra.  Dots are for the Tenti S6 model,
open balls are for the Tenti S7 model.  Full line: prediction using
Eq.\ (\ref{eq:relaxation}) with two rotational degrees of freedom and
relaxation time $\tau_r = 3.8\times 10^{-10}\: {\rm s}$.  The dashed line
represents the prediction of Eq.\ 28 in \cite{meador95}.
The points at $p = 1.5\;{\rm bar}$ (indicated with filled squares for an analysis
with Tenti S6 and open squares for Tenti S7)
were measured using 403.0~nm laser
light, and an FP-analyzer with $f_{\rm FSR} = 7553$~MHz, and $f_{\rm w} =
139$~MHz.
To avoid congestion, only the error bars for the S6 model are shown,
those of the S7 model are roughly the same.
}\label{fig.results}
\end{figure*}

The spectral response $S(f)$ of the FPI was
measured in a separate experiment, and could be parametrized very
well by the formula
\be
   S(f) = \left\{ 1 +
   [(2 f_{\rm FSR} / \pi f_{\rm w})
   \sin (\pi f / f_{\rm FSR})]^2 \right\}^{-1},
\label{eq.fp}
\ee
where $f_{\rm FSR}$ is the free spectral range of the FPI, $f_{\rm
FSR} = 7440$~MHz, and $f_{\rm w} = 232$~MHz is the Airy-width of the
transmission peak.  All computed model spectra were convolved with
$S(f)$, and since the free spectral range is relatively small, it is
important to allow for the periodic nature of $S(f)$.
The light that passes through the FPI is detected using a
photo-multiplier tube (PMT) which is operated in the photon-counting
mode and read out by a computer.

The experimental and computed spectra were normalized such that
$\int_{-f_b}^{f_b} I(f) \: \d f = 1$, where the integral extends over
one free spectral range (FSR), $f_b = f_{\rm FSR} / 2$.
In addition, the background of the model spectra was fitted to the
experimental data.  An estimate of the $\chi^2$ error was obtained
assuming Poissonian statistics of the photon counts.

The results are shown in Fig.\ \ref{fig.results}.  We fit the bulk
viscosity $\eta_b$ in both Tenti S6 and S7 models and find the values for
$\eta_b$ which minimize the $\chi^2$ difference between model and
experiment.
%
As Fig.\ \ref{fig.results}(e) indicates, the minimum of $\chi^2$ is
well defined at high pressures where the Brillouin side peaks
are pronounced, and not very well defined at $p =
1$~bar.  We find significant systematic differences between the model
and the experiment, corresponding to large values of $\chi^2$.
These differences are also shown in Fig.\ \ref{fig.results}(a--d).
The model-Brillouin peaks appear shifted towards larger (absolute)
frequencies compared to those of the experiment. The position of the
Brillouin peaks represents the velocity of sound, which is determined
by the internal degrees of motion of the molecule.  It is tempting to
vary the heat capacity of internal motion in order to obtain a better
fit.
At these frequencies only rotations should partake in the relaxation
of the internal energy, with the heat capacity of internal motion
$c_{\rm int} = 1$.  A slightly better fit of the peak locations could
be obtained by setting $c_{\rm int} = 1.16$, but now discrepancies at
other frequencies become more obvious.  Therefore, we kept $c_{\rm
int} = 1$, whilst we used $\eta = 1.46\times 10^{-5} \: {\rm kg \:
m}^{-1} {\rm s}^{-1}$ for the shear viscosity and $\kappa =
1.31\times10^{-2} \:{\rm W}\:{\rm K}^{-1}{\rm m}^{-1}$ for the
thermal conductivity \cite{pan.co2}.

The measured bulk viscosities are shown in Fig.\
\ref{fig.results}(f), which collects the results of two experimental
runs recorded at $\lambda =366.8$ nm, taken a few months apart
and an additional measurement recorded at $\lambda=403.0$ nm and $p=1.5$ bar.
The uncertainties in the derived
values for $\eta_{b}$ is composed of three contributions. The
typical contribution of the statistical uncertainty, as a result of
the fitting procedure shown in Fig.\ \ref{fig.results}(e), is $5
\times 10^{-8}\; {\rm kg} \: {\rm m}^{-1} \: {\rm s}^{-1}$. The
$0.9^\circ$ uncertainty in the determination of the scattering angle
translates into a systematic uncertainty of typically $5 \times
10^{-7} \:{\rm kg}\: {\rm m}^{-1} \: {\rm s}^{-1}$, while the 1\%
uncertainty in the pressure reading corresponds to a contribution of
$2 \times 10^{-7} \: {\rm kg}\: {\rm m}^{-1} \: {\rm s}^{-1}$. Here it
is noted that for the lower pressures ($p \leq 2$ bar), where the
information content of the spectrum is lower, all three contributions
to the uncertainty are larger. At the highest pressure ($p = 4$ bar),
where the effect of the bulk viscosity is most decisive, the total
uncertainty is less than $5 \times 10^{-7}\: {\rm kg}\: {\rm
m}^{-1}\:{\rm s}^{-1}$.  Further it should be noted that the
systematic uncertainty of the determination of the scattering angle
yields the largest contribution to the uncertainty budget, and that
all measurements were performed in the same scattering geometry.
Hence the relative uncertainties are lower than the error bars
indicated in Fig.\ \ref{fig.results}(f).

For large pressures in the range $p=2-4$ bar
we obtain $\eta_b = (6.0 \pm 0.3) \times 10^{-6} \:
{\rm kg \: m}^{-1} {\rm s}^{-1}$ using the Tenti S7 model
and $\eta_b = (4.5 \pm 0.6) \times 10^{-6} \:
{\rm kg \: m}^{-1} {\rm s}^{-1}$.  The measured $\eta_b$ appear to
{\em decrease} with increasing pressures, which does not agree with
the simple idea that at finite frequencies $\omega$, $\eta_b$ should
{\em increase} with increasing pressures, an idea which is embodied
by Eq.\ (\ref{eq:relaxation}).
We compare the measured pressure dependence of $\eta_b$ to the
predictions of Eq.\ (\ref{eq:relaxation}) and to Eq.\ 28 of
\cite{meador95} using a rotational relaxation time $\tau_r =
3.8\times 10^{-10}\: {\rm s}$.  These predictions disagree
significantly with the experiments.

An averaged value of $\eta_b = (5.7 \pm 0.6) \times 10^{-6}$~kg~m$^{-1}$s$^{-1}$
for the bulk viscosity as obtained via the Tenti S6 and S7
models in the pressure range $p=2-4$ bar can be compared to
$\eta_b = 4.6 \times 10^{-6}$~kg~m$^{-1}$s$^{-1}$ by Lao {\it et al.}
\cite{lao.1976a}, and
$\eta_b = (5.8\pm 1) \times 10^{-6}$~kg~m$^{-1}$s$^{-1}$ by Meijer
{\it et al.} using \crbs\ \cite{ru.jcp.2010}, but which is
somewhat larger than the value
$\eta_b = 3.7\times 10^{-6}$~kg~m$^{-1}$s$^{-1}$ found by Pan {\it et
al.} \cite{pan.co2}.
It is very different for light scattering experiments compared to
acoustical experiments performed at MHz frequencies.
A problem is the significant difference between experiment and the
Tenti models that were used to determine $\eta_b$.  For our
experiments on nitrogen gas at comparable pressures the Tenti S6
model fits the data much better \cite{all.rbs,vu.n2}.

The core part of the code that computes the Tenti models has been kindly provided to us by X. Pan. Also, the authors would like to thank A. G. Straume and O. Le Rille (European Space Agency), and B. Witschas (DLR Oberpfaffenhofen, Germany) for helpful discussions.
This work was funded by the European Space Agency, contract no. 21396.


\newpage

\section*{Informational Fifth Page}
In this section, please provide full versions of citations to
assist reviewers and editors (OL publishes a short form of
citations) or any other information that would aid the peer-review
process.

\end{document}